\documentclass[10pt,aps,prl,amsmath,amssymb,twocolumn,
				showpacs,reprint]{revtex4-1}

\usepackage{graphicx}% Include figure files
\usepackage{dcolumn}% Align table columns on decimal point
\usepackage{bm}% bold math

\begin{document}

\title{Ferroelectricity at ferroelectric domain walls}

\author{Jacek C. Wojde{\l}}
%\email{jcwojdel@icmab.es}
\affiliation{Institut de Ci\`encia de Materials de Barcelona
  (ICMAB-CSIC), Campus UAB, 08193 Bellaterra, Spain}

\author{Jorge \'I\~niguez}
%\email{jiniguez@icmab.es}
\affiliation{Institut de Ci\`encia de Materials de Barcelona
  (ICMAB-CSIC), Campus UAB, 08193 Bellaterra, Spain}

\begin{abstract}
We present a first-principles study of model domain walls (DWs) in
prototypic ferroelectric PbTiO$_{3}$. At high temperature the DW
structure is somewhat trivial, with atoms occupying high-symmetry
positions. However, upon cooling the DW undergoes a symmetry-breaking
transition characterized by a giant dielectric anomaly and the onset
of a large and switchable polarization. Our results thus corroborate
previous arguments for the occurrence of ferroic orders at structural
DWs, providing a detailed atomistic picture of a temperature-driven
DW-confined transformation. Beyond its relevance to the field of
ferroelectrics, our results highlight the interest of these DWs in the
broader areas of low-dimensional physics and phase transitions in
strongly-fluctuating systems.
\end{abstract}


% 77.80.-e Ferroelectricity and antiferroelectricity
% 77.80.B- Phase transitions and Curie point (for Curie point in
%          ferromagnetic materials, see 75.30.Kz):

% 63.70.+h Statistical mechanics of lattice vibrations and displacive phase
%         transitions

% 71.  Electronic structure of bulk materials
% 71.15.Mb Density functional theory, local density approximation,
% gradient and other corrections 

\maketitle

The structural domain walls (DWs) occurring in ferroelectric (FE) and
ferroelastic (FS) materials have become a focus of attention. Recent
studies show that the DWs can present a variety of properties, from
conductive \cite{seidel09,guyonnet11,farokhipoor11,aird98} and optical
\cite{yang10,alexe11} to magnetic \cite{he12,daraktchiev10,gareeva10},
that differ from those of the neighboring domains, which suggests that
they could be the active element in nano-technological applications
\cite{salje10,catalan12}. Elucidating the DW behavior poses major
experimental challenges, and the origin of most of the newly
discovered effects remains unclear. In fact, we still lack a detailed
structural and dynamical picture of the DWs, and in many cases we can
only speculate about the structure--property relationships at work
within them. Hence, there is a pressing need for predictive
theoretical studies tackling the DWs at an atomistic level and at the
relevant conditions of temperature, etc.

The DW structure, and even the possible occurrence of DW-confined
ferroic orders, have been discussed theoretically for decades, usually
in the framework of continuum Ginzburg-Landau or phenomenological
model theories
\cite{strukov-book1998,tagantsev-book2010,lajzerowicz79,houchmandzadeh91,tagantsev01,goncalves-ferreira08,lee09,conti11,stepkova12,taherinejad12,marton13}. Materials
with competing structural instabilities have been a focus of
attention, a good example being perovskite SrTiO$_{3}$ (STO). STO
undergoes a FS transition driven by an anti-ferrodistortive (AFD) mode
that involves concerted rotations of the O$_{6}$ octahedra in the
perovskite structure. This mode competes with a FE instability that is
suppressed by the onset of the AFD distortion \cite{zhong95b}. Yet,
there are both theoretical and experimental indications that a polar
order occurs at low temperatures {\sl within STO's FS DWs}
\cite{tagantsev01,scott12,eliseev12,salje13}, i.e., in the region
where the otherwise dominant AFD distortions vanish. In this context,
it is worth noting recent first-principles studies predicting that
PbTiO$_{3}$ (PTO) \cite{wojdel13} and related compounds
\cite{kornev06} present a FE-AFD competition that is even stronger
than the one occurring in STO. These are the ideal conditions to
obtain interesting effects at structural DWs, and motivated this work.

\begin{figure*}[t!]
\includegraphics[width=0.95\linewidth]{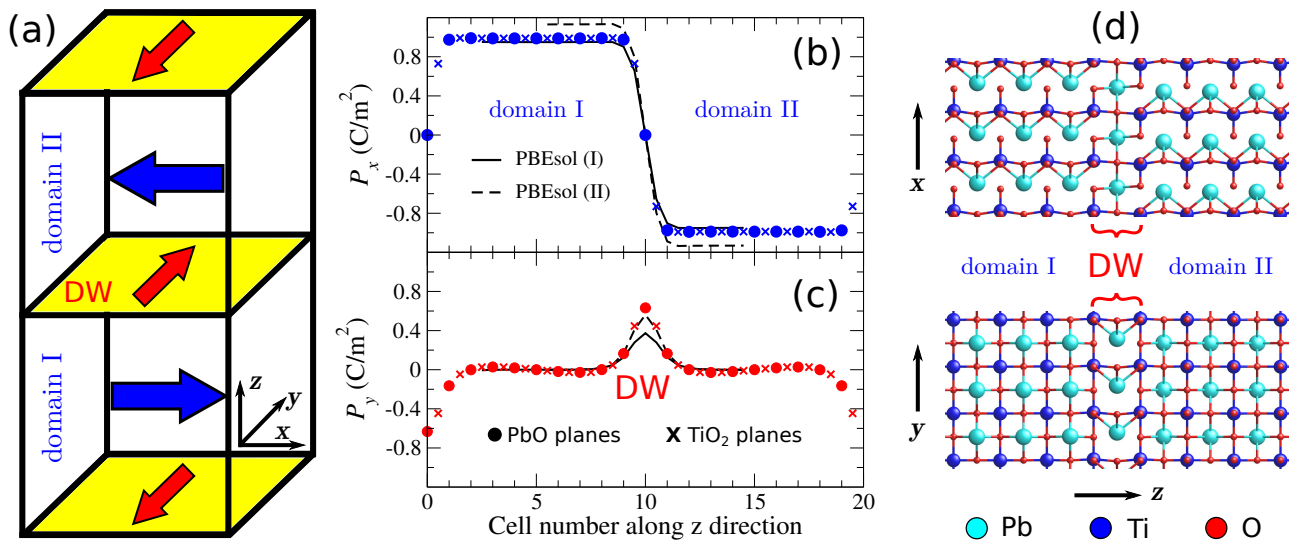}
\caption{Panel~(a): Sketch of the supercell used in our
  simulations. The indicated Cartesian axes coincide with the
  principal directions of the perovskite lattice. Panels~(b) and (c):
  Polarization profile corresponding to the stable structure of our
  multi-domain configuration. (Calculation of local polarizations
  described in Ref.~\protect\cite{supp}.) The ``PBEsol (I)'' lines
  show the results of an unconstrained first-principles structural
  relaxation \protect\cite{supp}; the agreement with the model
  predictions is essentially perfect for the $P_{x}$ profile; for the
  DW polarization we obtain a slightly smaller value. The ``PBEsol
  (II)'' lines show the results of a first-principles relaxation in
  which the supercell lattice parameters were fixed to match those
  predicted by our model potential; the agreement for the $P_{y}$
  profile is essentially perfect; for the polarization within the
  domains we get slightly larger values. Panel~(d): Views of the
  atomic structure of our multi-domain configuration.}
\label{fig1}
\end{figure*}

{\sl Low-temperature study}.-- We employed the tools of
Ref.~\onlinecite{wojdel13}, which permit large-scale simulations with
first-principles predictive power, to investigate an ideal version of
the simplest DWs occurring in PTO, namely, 180$^{\circ}$ boundaries
separating regions of opposed polarization and being perfectly
planar. We used the model potential for PTO labeled ``$L^{I}$''
\cite{wojdel13}, which we briefly describe in \cite{supp}. As shown in
Fig.~\ref{fig1}(a), we set the polarization of the first domain ${\bf
  P}^{\rm I}$ parallel to the [100] direction of the perovskite
lattice, and took ${\bf P}^{\rm II}\parallel [\bar{1}00]$ for the
second one; the DW in between was assumed to reside in a (001)
plane. Our supercell, which contains 12$\times$12$\times$20 perovskite
units (14400 atoms), is periodically repeated and holds two DWs.

We investigated the ground state structure of this multi-domain
configuration by means of Monte Carlo (MC) simulated annealings
\cite{supp}. Figure~\ref{fig1}(b) shows the $x$-component of the
polarization ($P_{x}$) as we move along $z$. We observe two domains
within which PTO adopts the structure of its homogeneous ground state,
with an associated polarization of about 0.99~C/m$^{2}$ and a cell
aspect ratio of about 1.07. The domains are separated by a DW centered
at a PbO plane and presenting a thickness of about one unit cell.

Our DWs do not display any rotations of the O$_{6}$ octahedra. This
result lends itself to a simple explanation: Because the DWs are
ultra-thin, hypothetical DW-localized AFD modes would {\em overlap}
with the neighboring FE distortions, and thus be penalized by the
FE-AFD competition. As a result, the absence of localized AFD modes
seems rather natural.

\begin{figure}[t!]
\includegraphics[width=0.95\linewidth]{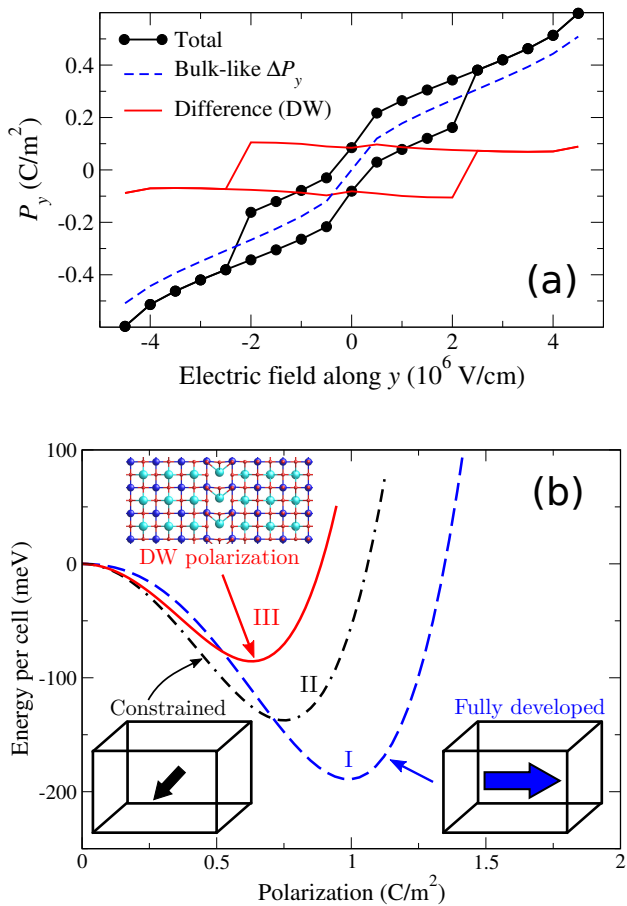}
\caption{Panel~(a): Polarization loop computed for the $x$-polarized
  multi-domain state of Fig.~\protect\ref{fig1} and for an electric
  field along the $y$ direction. For these simulations we used a
  1$\times$1$\times$20 supercell, thus assuming an homogeneous switch,
  and performed the calculations in the limit of 0~K (details in
  \cite{supp}). The total response (black line and circles) is split
  in two parts: (1) The response of an $x$-polarized mono-domain state
  (dashed blue line); note that this response does not saturate, as
  the $P_{x}$ polarization will eventually rotate to align with the
  field. (2) The difference between the total and the mono-domain
  results (solid red line), which captures the DW response. In order
  to observe the switch at the DWs, we imposed in the simulations the
  strain of the multi-domain ground state, and thus prevented the
  domain polarizations from fully aligning with the applied field for
  relatively small field values. Such a constraint is similar to the
  epitaxial one characteristic of thin films. Panel~(b): Energy wells
  corresponding to FE instabilities in various situations described in
  the text. Note that the $P = 0$ state corresponds to a different
  atomistic configuration in each case. In the case of the DW
  polarization (red solid line) the energy is given {\sl per} cell
  within the DW plane.}
\label{fig2}
\end{figure}

Nevertheless, the structure of the DWs is far from being trivial. As
shown in Fig.~\ref{fig1}(c), a non-zero $P_{y}$ polarization appears
at the DW plane and rapidly vanishes as we move into the domains. This
DW polarization is switchable, as evidenced by the hysteresis loop in
Fig.~\ref{fig2}(a). Further calculations show that the polarizations
of neighboring DWs couple, and tend to align in a parallel
configuration when the walls are sufficiently close. However, the
DW--DW interaction quickly decreases with the separation distance; for
example, for our 12$\times$12$\times$20 supercell, the energy split
between the parallel and anti-parallel states is about 0.01~meV {\sl
  per} DW cell, which is negligible. Hence, the anti-parallel
configuration shown in Fig.~\ref{fig1} is a stable state (i.e., a
local minimum of the energy), but not more significant than the also
stable, quasi-degenerate parallel configuration.

Our results show how strain and the reduced dimensionality determine
the energetics of the DW-confined FE distortion. The onset of the
multi-domain structure for $P_{x}$ implies a strong deformation of the
perovskite lattice: it becomes tetragonal and acquires an aspect ratio
of 1.07, the long lattice vector coinciding with the polar axis
$x$. In the case of our simulated system, the $xy$ plane is
homogeneously strained throughout the supercell (stretched along $x$,
shrunk along $y$). Hence, even if we have $P_{x} = 0$ at the DWs of
Fig.~\ref{fig1}, the strain disfavors the occurrence of a DW
polarization along the $y$ direction, which is subject to a
compression. The ensuing effect can be appreciated in the potential
wells of Fig.~\ref{fig2}(b): Case~I corresponds to the full
development of the FE distortion of PTO, as it happens within the
domains. Case~II corresponds to the development of a three-dimensional
$y$-polarized state when we constrain the cell to be strained as in
the $x$-polarized FE state; the equilibrium polarization and
associated energy gain get clearly reduced, and we obtain a value of
$P_{y}$ (about 0.75~C/m$^{2}$) that is not far from our result at the
DW center (about 0.65~C/m$^{2}$). Additionally, Fig.~\ref{fig2}(b)
shows a case~III corresponding to the condensation of $P_{y}$ at our
DWs; the energy well becomes shallower than in case~II, indicating a
further weakening of the polar instability caused by the spatial
confinement (i.e., by the truncation of interactions favoring the
three-dimensional homogeneous polar state) and the competition with
the $P_{x}$ distortion of the neighboring domains. Nevertheless, the
obtained well depth (86~meV/cell) is sizable, which suggests that the
predicted DW instability should occur at relatively high temperatures.

We checked the correctness of our model-potential predictions by
running direct first-principles calculations of our multi-domain
structure, using a 1$\times$1$\times$20 cell. As shown in
Figs.~\ref{fig1}(b) and \ref{fig1}(c), the agreement between our
model-potential and first-principles results is very good, and the FE
character of PTO's DWs is confirmed \cite{fncell}. We should note that
there are several first-principles studies of the 180$^{\circ}$ DWs of
PTO in the literature~\cite{poykko99,meyer02,he03,lee09}, and the
consensus is that no DW-confined polarization occurs. We cannot be
sure about the reasons why these previous works did not find polarized
DWs; some possibilities are discussed in \cite{supp}.

We checked whether this confined polarization occurs in other PTO
DWs. We found that 180$^{\circ}$ DWs lying in other planes -- e.g., a
(011) boundary separating domains with ${\bf P}^{\rm I} \parallel
[100]$ and ${\bf P}^{\rm II} \parallel [\bar{1}00]$ -- present polar
distortions analogous to the one just described. In contrast, we found
that 90$^{\circ}$ DWs do not present any FE instability, a result
probably related with the fact that these boundaries are considerably
more distorted than their 180$^{\circ}$ counterparts, or to the
elastic (epitaxial tensile) constraints we had to impose in order to
stabilize them. PTO's 90$^{\circ}$ DWs will be discussed elsewhere.

Finally, let us note that the low-temperature configuration of our PTO
DWs can be described as being Bloch-like. First-principles theory has
predicted the occurence of Bloch-like DW configurations in materials
like LiNbO$_{3}$ \cite{lee09} and BaTiO$_{3}$ in its rhombohedral
phase \cite{taherinejad12}.

\begin{figure}[t!]
\includegraphics[width=0.95\linewidth]{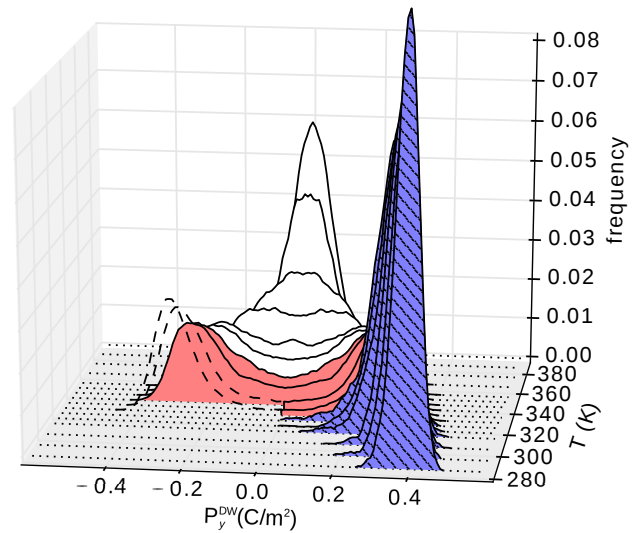}
\caption{Probability distribution $\rho(P_{y}^{\rm DW})$ for the
  $y$-component of the polarization at the center of the DW and as a
  function of temperature. The three temperature regions mentioned in
  the text are marked with different colors: white for $T \geq 350$~K
  (where we clearly have $\langle P_{y}^{\rm DW} \rangle = 0$), red
  for 320~K~$< T <$~350~K (critical region), and blue with stripes for
  $T \leq 320$~K (where we clearly have $\langle P_{y}^{\rm DW}
  \rangle \neq 0$). The left side of the histograms for $T = 330$~K
  and $T = 325$~K is drawn using dashed black lines; this is to
  emphasize that, at these temperatures, we attribute the $P_{y}^{\rm
    DW}$ fluctuations from positive to negative values to finite size
  effects (see \cite{supp}).}
\label{fig3}
\end{figure}

{\sl Behavior with temperature}.-- We studied PTO DWs as a function of
increasing temperature by running MC simulations as described in
\cite{supp}. Figure~\ref{fig3} shows the obtained probability density,
$\rho(P_{y}^{\rm DW})$, for the $y$-component of the polarization at
the DW plane. We observe three distinct regions: (1) For $T \leq
320$~K the DW presents a stable and large polarization, and the
equilibrium state resembles the one discussed above. (2) For $T \geq
350$~K the DW gets disordered and we have a null thermal average
$\langle P_{y}^{\rm DW}\rangle = 0$. At those temperatures the system
presents a mirror symmetry plane perpendicular to the $y$ axis, and we
could say that the DWs are in a {\em paraelectric} state
\cite{fnsymm}. (3) Finally, for 320~K~$< T <$~350~K we have a narrow
region in which, during the course of the MC simulation, $P_{y}^{\rm
  DW}$ occasionally switches between equivalent polar states. Clearly,
the finite size of our simulation supercell is partly responsible for
such fluctuations, which should be considered a spurious finite size
effect below a certain transition temperature $T_{\rm C}^{\rm DW}$. At
any rate, the presence of a phase transition is obvious from these
results, and the analysis of the MC data suggests that we have $T_{\rm
  C}^{\rm DW} \approx 335$~K. (See \cite{supp} for details.)

\begin{figure}[t!]
\includegraphics[width=0.95\linewidth]{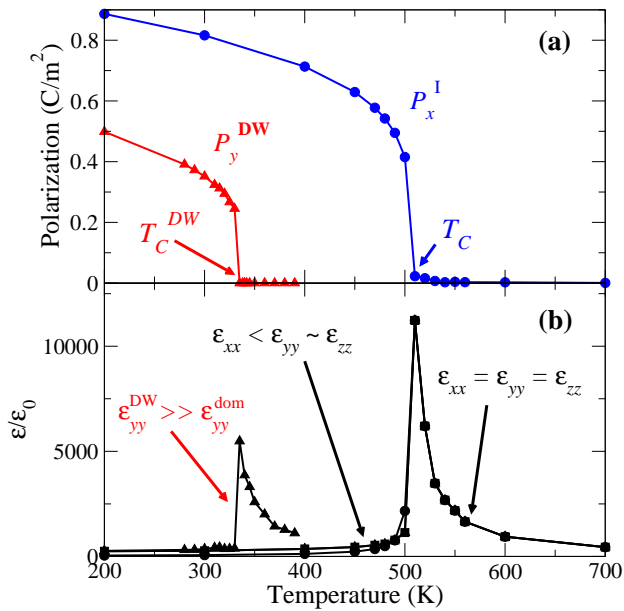}
\caption{Panel~(a): Temperature dependent polarization at the center
  of the first domain ($P_{x}^{\rm I}$, blue circles) and at the
  center of the DWs ($P_{y}^{\rm DW}$, red triangles). Panel~(b):
  Diagonal components of the dielectric permittivity tensor. The
  discontinuity for $\epsilon_{yy}$ (triangles) at $T = 390$~K is
  related with the finite-size effects described in
  \cite{supp}.}
\label{fig4}
\end{figure}

The phase transitions in our simulated system are clearly appreciated
in Fig.~\ref{fig4}, which shows the evolution of the relevant order
parameters and dielectric response. As the system cools down from high
$T$, the diagonal components of the dielectric tensor increase
sharply, revealing a FE transition at $T_{\rm C} = 510$~K. [The
  quantitative disagreement between our computed $T_{\rm C}$ and the
  experimental result for PTO (760~K) is discussed in
  Ref.~\onlinecite{wojdel13}.] Note that for $T > 510$~K we have
$\epsilon_{xx} = \epsilon_{yy} = \epsilon_{zz}$, reflecting the cubic
symmetry of the paraelectric phase. Then, for $T \gtrsim T_{\rm
  C}^{\rm DW}$ the $\epsilon_{yy}$ component [triangles in
  Fig.~\ref{fig4}(b)] behaves in an anomalous way. As regards the
dielectric response along the $y$ direction, our multi-domain
configuration can be viewed as a set of parallel capacitors, so that
the total dielectric constant can be split into domain and DW
contributions, $\epsilon_{yy} = f^{\rm dom} \epsilon_{yy}^{\rm dom} +
f^{\rm DW} \epsilon_{yy}^{\rm DW}$, where $f^{\rm dom}$ and $f^{\rm
  DW}$ are the respective volume fractions. Noting that $f^{\rm dom}
\approx 0.9$ and $f^{\rm DW} \approx 0.1$ in our case, and that
$\epsilon_{yy}^{\rm dom}$ is featureless for $T < T_{\rm C}$ (which we
checked by running MC simulations of the mono-domain case), it is
clear that the increase in $\epsilon_{yy}$ comes from a very large DW
response $\epsilon_{yy}^{\rm DW}$. Hence, the DW-confined transition
is driven by a FE soft mode and characterized by very strong
fluctuations of the order parameter, as consistent with the reduced
dimensionality.

{\sl Final remarks}.-- As mentioned, the possibility of having polar
orders and dielectric anomalies associated with DWs has been discussed
in the theoretical literature, typically employing phenomenological
continuum models \cite{tagantsev01,marton13}. Our first-principles
work corroborates that such effects can indeed occur, providing for
the first time a realistic atomistic picture of a $T$-driven
DW-confined ferroic transformation.

One might describe the found transition as a change in the DW
character, from Bloch to Ising, upon heating; in fact, some authors
have discussed similar effects in these terms
\cite{stepkova12,marton13}. However, we think that our result is
better described as a proper FE phase transition confined to the DW,
to emphasize that it results in a switchable DW polarization. For the
same reasons, we would rather denote the low-temperature state as a
ferroelectric DW, and not simply as a DW with Bloch-like
character. Note that, as in the case of LiNbO$_{3}$ \cite{lee09},
Bloch-like DWs may not display a net polarization.

The present results are the first step in the investigation of such an
interesting phenomenon. Aspects for future work include: the possible
critical behavior of the transition, the pre-transitional dynamics,
the internal structure of the DWs (can we have multi-domain states
within them?), and the role of DW--DW interactions (can it affect the
dimensionality and features of the transition?). We thus believe our
findings will open exciting research avenues in the fields of phase
transitions and low-dimensional physics.

It may seem surprising that the predicted effect has not been reported
experimentally. However, note that observing such a FE order may
require some uncommon measurements. Ideally, one would like to work
with samples presenting a pattern of highly-ordered stripe domains
separated by 180$^{\circ}$ DWs, as displayed by suitably grown PTO
films \cite{streiffer02} and PTO/STO superlattices
\cite{zubko10,zubko12}. High-resolution X-ray measurements at low
temperatures may reveal the DW order. Additionally, by measuring the
dielectric response along the in-plane direction of the DWs, one
should be able to observe a clear feature around $T_{\rm C}^{\rm
  DW}$. We hope our results will motivate further experimental work to
characterize FE DWs and the transitions that may occur within them.

\begin{acknowledgments}

Work supported by MINECO-Spain (Grants No. MAT2010-18113 and
No. CSD2007-00041) and CSIC [JAE-doc program (JCW)]. Some figures were
prepared using VESTA~\protect\cite{vesta}. We thank Ekhard Salje,
David Vanderbilt, and Pavlo Zubko for their useful comments.

\end{acknowledgments}

\section{\bf Supplemental Information for ``Ferroelectric transitions at
  ferroelectric domain walls''}

In the following we describe the simulation methods employed as well
as various technicalities pertaining to the calculation of some
quantities. We also comment on the determination and quantitative
accuracy of the temperature at which the domain wall-confined
ferroelectric transition takes place, and on previous first-principles
literature on the DWs in PbTiO$_{3}$.

\section{Methods}

For most simulations we used the first-principles model for PTO
described in Ref.~\onlinecite{wojdel13}, where it is labeled
``$L^{I}$''. This model can be viewed as a Taylor series of the
energy, around the ideal cubic perovskite structure, as a function of
all possible atomic distortions and strains. The series was truncated
at 4th order and only pairwise interaction terms were included. The
potential parameters were computed by using the local density
approximation (LDA) to density functional theory (DFT). To compensate
for LDA's well-known overbinding problem, we simulate the model under
the action of a tensile hydrostatic pressure of 14.9~GPa.

We ran some direct first-principles simulations to verify the model
predictions at low temperatures. For that task, we used the
PBEsol\cite{perdew08} approach to DFT as implemented in the VASP
package.\cite{kresse96} Note that the PBEsol functional has been shown
to render accurate results for the equilibrium structures of several
FE perovskite oxides\cite{wahl08} including PTO (which we checked
ourselves). We used the “projector augmented wave” method to represent
the ionic cores,\cite{blochl94,kresse99} solving for the following
electrons: Ti's 3$p$, 3$d$, and 4$s$; Pb's 5$d$, 6$s$, and 6$p$; and
O's 2$s$ and 2$p$. Wave functions were represented in a plane-wave
basis truncated at 500~eV, and a 6$\times$6$\times$1 $k$-point grid
was used for integrations within the Brillouin zone corresponding to a
1$\times$1$\times$20 supercell. To better compare our model-potential
and first-principles results, we conducted PBEsol structural
relaxations in both an unconstrained way [``PBEsol (I)'' results in
  Fig.~1 of our manuscript] and by imposing lattice vectors obtained
from model-potential calculations [``PBEsol (II)''].

For the sake of comparison, we also ran some simulations at the LDA
level, all other technicalities being identical to those of our PBEsol
calculations.

We computed the electric dipoles caused by the displacement of
individual atoms within a linear approximation, using the
corresponding dynamical charge tensors as computed for the cubic
phase. To compute Ti-centered local polarizations associated to
specific cells, as those shown in Fig.~1, we proceeded in the
following way: We summed up the dipoles associated to the central
titanium and to its nearest-neighboring atoms (i.e., 6 oxygens and 8
leads), weighting the contribution of the neighbors appropriately
(e.g., since an oxygen atom is shared by two Ti atoms, it contributes
one half of its associated dipole to the polarization centered at each
of its two neighboring titaniums). We proceeded analogously to compute
Pb-centered polarizations. This is a rather standard
scheme\cite{meyer02} related with the local-mode construction in
classic first-principles models of ferroelectrics.\cite{zhong95a}

Finally, we used a standard Monte Carlo (MC) scheme to account for the
effect of temperature in our system, working with a periodically
repeated box of 12$\times$12$\times$20 PTO unit cells. At all
temperatures we initialized our MC simulations from the multi-domain
ground state configuration described in Fig.~1. We typically ran
25,000~MC sweeps for thermalization, followed by at least 25,000
additional sweeps for computing thermal averages. In the region of the
DW-confined phase transition, marked in red in Fig.~3 of our
manuscript, we did production runs of up to 150,000~MC sweeps, as
needed to obtain acceptable statistics for $\rho(P_{y}^{\rm
  DW})$. Note that, because of the quasi-two-dimensional and
strongly-fluctuating nature of the polar order at the walls, this was
an especially challenging task; in particular, as discussed below, our
ability to get quantitatively accurate results in the critical region
is limited by the size of the considered supercell, which is the
largest we can afford with our current simulation tools. We also ran a
simulation of the FE phase transition of PTO, always in a mono-domain
configuration, using a 12$\times$12$\times$12 simulation box
periodically repeated. In this case, we always started the simulations
from a perfectly cubic structure; then we ran 10,000~MC sweeps for
thermalization, followed by 50,000 sweeps for computing averages,
which was enough to obtain well converged results at all
temperatures. We checked that the temperature dependent mono-domain
polarization thus computed is essentially identical to the local
polarization obtained at the middle of the domains in our multi-domain
simulations.

Whenever we needed to run a structural relaxation with our model
potential, we performed MC simulated annealings in which we reduced
the temperature down to essentially 0~K, allowing the simulation to
evolve until the system would get {\em frozen} at an energy
minimum. In some cases we started the annealings at very low
temperatures, which allowed us to investigate local (meta-stable)
minima of the energy, as the lack of thermal activation prevents the
simulated system from evolving to its most stable configuration. This
was important, for example, to compute the hysteresis loops of
Fig.~2(a) of our manuscript, as we needed to follow the evolution of
the polar states, as a function of the magnitude of an opposing
electric field, up to their meta-stability limit. Beyond that limit,
the local energy minimum is lost and the relaxation (i.e., the
simulated annealing) naturally evolves towards the polar state favored
by the applied field. Using the same strategy, we were also able to
quantify the energy gap between the states with parallel and
anti-parallel DW polarizations, and thus determine that the DW--DW
interaction is negligible for the separating distance (i.e., 9 unit
cells or about 35~\AA) corresponding to our simulation box.

\begin{figure}[t!]
\includegraphics[width=0.95\columnwidth]{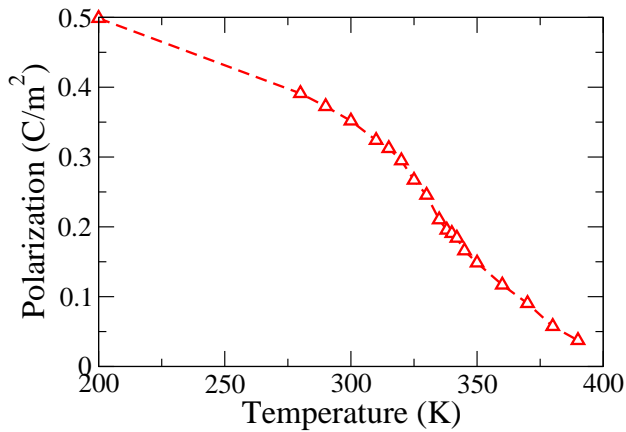}
\caption{Computed temperature dependence of $\langle |P_{y}^{\rm DW}|
  \rangle$, the absolute value of the $y$ component of the DW
  polarization. From the inflection point of this curve, obtained
  numerically, we can estimate $T_{\rm C}^{\rm DW} \approx 332$~K.}
\label{fig1}
\end{figure}

\begin{figure*}[t!]
\includegraphics[width=0.95\linewidth]{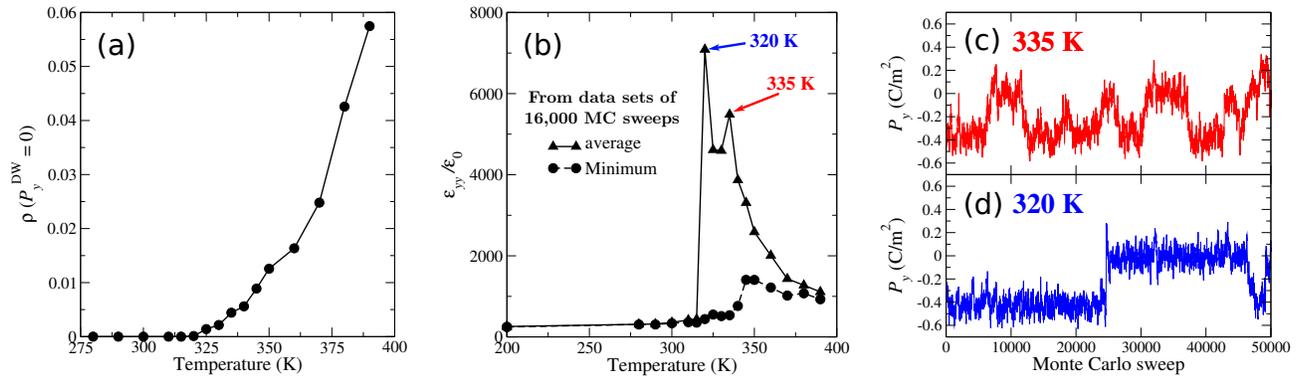}
\caption{Panel~(a): Probability $\rho(0)$ of having $P_{y}^{\rm DW} =
  0$ as a function of temperature. Panel~(b): $\epsilon_{yy}$
  component of the dielectric permittivity tensor, as a function of
  temperature, and estimated in the two different ways described in
  the text. Panels~(c) and (d): Evolution of $P_{y}$ during 50,000
  sweeps of the MC runs at 335~K and 320~K, respectively. The shown
  data is representative of the behavior of our simulated system
  during the longer runs.}
\label{fig1}
\end{figure*}

\section{Finite size effects in the Monte Carlo simulations}

As already mentioned in the manuscript, at temperatures around the
DW-confined transition our MC results are affected by the fact that
our simulation supercell is finite in size. More specifically, we
simulate DWs that are composed of 12$\times$12 cells periodically
repeated in the DW plane. While we have checked that a lateral size of
12~cells is sufficient to describe the bulk FE transition of PTO with
good accuracy, our results suggest that we may not be converged
size-wise as regards the DW-confined transformation (see Fig.~3 of our
manuscript). Whenever one deals with low-dimensional orders, large
amplitude fluctuations are more likely to occur, which complicates
convergence with size. Unfortunately, at present our software does not
allow us to further increase the size of the simulation box, as the
calculations become too heavy computationally. Nevertheless, an
analysis of the results obtained for 12$\times$12 DWs allows us to
bracket $T_{\rm C}^{\rm DW}$ quite precisely.

First of all, let us stress that the uncertainty caused by these size
effects is of little practical importance. The affected temperature
range is relatively small, approximately from 320~K to 350~K. Hence,
from the point of view of comparing our value of $T_{\rm C}^{\rm DW}$
with an eventual experimental determination, the error coming from
size effects can be considered negligible.

There are simple and usually effective strategies to estimate the
temperature dependence of the order parameter, from our MC data, in
ways that are robust against size effects. Thus, for example, if we
are interested in the thermal average of the DW polarization $\langle
P_{y}^{\rm DW} \rangle$, it is a good idea to compute the average of
its absolute value, $\langle |P_{y}^{\rm DW}| \rangle$, instead. In
this way, spurious fluctuations (i.e., {\em jumps} between equivalent
energy wells) of the DW polarization, permitted by the finite size of
the simulation box, will not result in a vanishing order parameter. An
obvious drawback of this approach is that $\langle |P_{y}^{\rm DW}|
\rangle$ will be different from zero even in the disordered phase,
which is not correct. Yet, monitoring the temperature dependence of
this quantity is usually enough to locate a phase transition, and even
permits a quantitative estimation of the transition temperature.

Figure~S1 shows the results for $\langle |P_{y}^{\rm DW}| \rangle$
from our MC simulations. As expected, we obtain a smooth temperature
dependence that clearly points at a transition occurring in the range
between 320~K and 350~K. Further, we can estimate the transition
temperature by numerically calculating the inflection point of this
curve, which happens at about 332~K.

We were not fully satisfied by this simple estimate, and tried to
analyze our MC data in more detail, proceeding as
follows. Figure~S2(a) shows the computed $\rho(P_{y}^{\rm DW})$
evaluated at $P_{y}^{\rm DW} = 0$ and as a function of
temperature. There we can appreciate that at $T \leq 320$~K the DWs do
not visit the $P_{y}^{\rm DW} = 0$ state, as we have $\rho(0) \approx
0$. Hence, it is clear that our simulations predict $T_{\rm C}^{\rm
  DW} > 320$~K. Above that temperature, there is a range in which we
get very small but non-zero values of $\rho(0)$, as well as a bimodal
shape for $\rho(P_{y}^{\rm DW})$. Do we really have disordered DWs at
these temperatures?

To gain insight into the behavior of the system in this temperature
range, we considered several ways to compute the dielectric constant
$\epsilon_{yy}$. Let us recall that $\epsilon_{yy}= \epsilon_{0}
(\chi_{yy} +1)$, and that we have
\begin{equation}
\epsilon_{0} \chi_{yy} = \frac{\partial P_{y}}{\partial E_{y}} =
\frac{\beta}{V} (\langle P_{y}^{2} \rangle - \langle P_{y}
\rangle^{2}) \, ,
\label{eq:chi}
\end{equation}
where $E_{y}$ is the applied electric field, $V$ is the supercell
volume, $\beta = 1/k_{\rm B}T$, $k_{\rm B}$ is the Boltzmann constant,
and $\langle ... \rangle$ denotes a thermal average as computed from
our MC simulations. Now, instead of using all the sweeps of our
production MC runs to compute $\chi_{yy}$, it was instructive to
consider smaller sets of data that we label by $i$, compute the
corresponding $\chi_{yy}^{(i)}$ for each set, and perform a
statistical analysis of the results. Figure~S2(b) summarizes the
outcome of such an exercise. We show the results obtained by
considering sets of data composed of 16,000~MC sweeps each. (We
checked that 16,000~sweeps are typically sufficient to get
well-converged thermal averages of our quantities of interests.) If we
approximate $\chi_{yy}$ by
\begin{equation} 
\bar{\chi}_{yy} = \frac{1}{M} \sum_{i=1}^{M} \chi_{yy}^{(i)} \, ,
\end{equation} 
where $M$ is the number of 16,000-sweep sets that we have, we obtain
the results shown by a solid black line and triangles in
Fig.~S2(b). Interestingly, $\bar{\chi}_{yy}$ presents two peaks
located at 320~K and 335~K, respectively. If, alternatively, we
approximate $\chi_{yy}$ by
\begin{equation}
\chi^{\rm min}_{yy} = \min_{i} \chi_{yy}^{(i)} \, ,
\end{equation}
we get the results shown by a dashed black line and circles in
Fig.~S2(b). In this case, we get a single broad peak with a maximum at
$T \approx 340$~K.

To understand the origin of these results, it is useful to look at the
evolution of $P_{y}$ during the MC runs, as shown in Figs.~S2(c) and
S2(d) for 335~K and 320~K, respectively. At 335~K we observe frequent
fluctuations in the value of $P_{y}$, suggesting that the DWs are not
ordered. (Note that we have two DWs in the simulation cell; at 335~K
we get $|P_{y}|\approx 0.4$~C/m$^{2}$ when the two DW polarizations
are parallel, and $|P_{y}|\approx 0$ when they are anti-parallel.) The
fluctuations are large, which leads to a large response according to
Eq.~(\ref{eq:chi}) and explains the peak that $\bar{\chi}_{yy}$
presents at that temperature. At the same time, it is possible to find
relatively long segments of the MC run in which $P_{y}$ fluctuates
around a constant value. Hence, the computed $\chi_{yy}^{\rm min}$ is
much smaller than $\bar{\chi}_{yy}$.

Figure~S2(d) shows that the situation at 320~K is very different. In
this case, our DWs clearly present a stable polarization. Nevertheless,
from time to time we do observe a jump in $P_{y}$, indicating a switch
in the polarization of one of the DWs. These infrequent jumps lead to
a very large value of $\bar{\chi}_{yy}$, which peaks at that
temperature. In contrast, we find that $\chi_{yy}^{\rm min}$, which is
obviously determined by a segment of the MC trajectory that is free of
jumps in $P_{y}$, takes a relatively small value. From this analysis,
we can conclude that the peak observed at 320~K in $\bar{\chi}_{yy}$
is an artifact associated to the finite size of our simulation
supercell, and that our results suggest that $T_{\rm C}^{\rm DW}$ is
about 335~K or 340~K. To generate the figures in our manuscript, we
assumed $T_{\rm C}^{\rm DW} = 335$~K and processed our MC data
accordingly.

Finally, let us note another feature of our results that is related
with the finite size of our simulation box. In Fig.~4(b) of our
manuscript, the results for $\epsilon_{yy}$ with a clear DW
contribution extend only up to $T \approx 390$~K, where a
discontinuity can be appreciated. The reason is that above that
temperature, because of the finite size our supercell, the initial
multi-domain configuration evolves during the MC run into a
mono-domain state, which is more stable for the elastic
(unconstrained) and electrical (short-circuit) boundary conditions
considered in our simulations. Nevertheless, we do not observe any
sort of structural order at the DWs for $T \gtrsim T_{\rm C}^{\rm
  DW}$, and have no reason to expect a different behavior at higher
temperatures. Hence, our current inability to investigate the DWs up
to $T_{\rm C}$ is not troublesome.

\section{First-principles and model-potential accuracy}

The largest quantitative uncertainty affecting our value for $T_{\rm
  C}^{\rm DW}$ is surely related with the inaccuracies inherent to our
first-principles model potentials. As shown in
Ref.~\onlinecite{wojdel13}, our potential for PTO reproduces the
energetics of the FE instabilities with very good accuracy, and it
also accounts well for the lowest-energy perturbations of the relevant
stable structures. Hence, as far as the homogeneous FE phases of the
material are concerned, our potential can be considered an accurate
approximation to density functional theory (DFT) results obtained
using the local density approximation (LDA). However, the computed
$T_{\rm C} = 510$~K turns out to be significantly smaller than the
experimental value of 760~K, which suggests there is a problem with
the accuracy of the LDA and employed first-principles methods. As
discussed in Ref.~\onlinecite{wojdel13}, tracking down the origin of
this error is not trivial; the LDA could be giving us (1) FE
instabilities that are too weak, (2) AFD instabilities that are too
strong, (3) a FE-AFD competition that is too strong, or (4) a
combination of all these problems. All such errors will affect our
calculation of $T_{\rm C}^{\rm DW}$ in essentially the same way as
they affect the calculation of $T_{\rm C}$. Hence, we have reasons to
believe that our model may underestimate significantly the temperature
of the DW-confined transition.

Additionally, one may wonder whether our model reproduces the DFT
energetics associated with the DWs as well as it captures those
related with the bulk FE state. Our model renders a DW energy $E^{\rm
  DW} = 107$~mJ/m$^{2}$ when we allow for a full structural relaxation
of the DW, and $E^{\rm DW} = 190$~mJ/m$^{2}$ when we constrain the DW
to remain in the unpolarized high-symmetry state. Using the PBEsol
approach to DFT [which we denote ``PBEsol~(I)'' in our manuscript; see
  Methods section], we get 149~mJ/m$^{2}$ and 152~mJ/m$^{2}$,
respectively, for these two quantities. When we run the PBEsol
relaxation imposing the cell strain obtained from our model
calculations [``PBEsol~(II)''], we get 225~mJ/m$^{2}$ and
250~mJ/m$^{2}$, respectively. Note that Meyer and Vanderbilt reported
a value of 132~mJ/m$^{2}$ for this quantity,\cite{meyer02} which they
computed at the LDA level and considering an unpolarized DW. Thus, we
can conclude that our model gives acceptable results for the DW
energy, even though this piece of information was not used when
calculating the potential parameters. (For reference, we ran LDA
calculations considering an unpolarized DW, as done in
Ref.~\onlinecite{meyer02}, and got a DW energy of 97~mJ/m$^{2}$.)

For us it is more relevant to consider the energy difference between
the paraelectric and ferroelectric states of the DW, i.e., the depth
of the energy well corresponding to case~III in Fig.~2(b) of our
manuscript. For this quantity we get 86~meV/cell with our model,
4~meV/cell at the ``PBEsol~(I)'' level, and 26~meV/cell at the
``PBEsol~(II)'' level. It is interesting to note that, despite the
good agreement for the DW polarization and structure between our model
and the PBEsol calculations [see Fig.~1(c)], the differences in energy
are large. This comparison suggests that, with respect to the result
one would obtain from a PBEsol simulation, our model may overestimate
significantly the value of $T_{\rm C}^{\rm DW}$. [For reference, we
  also computed the depth of the energy well corresponding to the
  DW-confined FE instability at the LDA level, and obtained 6~meV/cell
  when a full structural relaxation is allowed, and 45~meV/cell when
  the cell obtained from our model-potential simulations is
  imposed. Hence, LDA renders slightly stronger DW-confined
  instabilities than PBEsol.]

How worrisome are the differences between our model and PBEsol
energetics as regards the DW order? How well would a PBEsol simulation
reproduce the $T_{\rm C}$ of the FE transition occurring in the bulk
of PTO? Is PBEsol the best DFT flavor available for this task?  It is
not possible to answer those questions at present. As an indication,
let us mention that our PBEsol calculations render a value of
67~meV/cell for the energy difference between the bulk-paraelectric
and bulk-ferroelectric phases of PTO. This value corresponds to the
depth of the energy well of case~I in Fig.~2(b), for which we get
189~meV/cell using our model. Hence, since our model underestimates
PTO's experimental $T_{\rm C}$ quite significantly, we can tentatively
conclude that a PBEsol simulation would lead to very small Curie
temperature for this material. Hence, despite its accuracy at
predicting equilibrium structures, it is not clear at all whether
PBEsol should be our DFT method of choice for calculating the
energetics of PTO's FE instabilities.

We hope this discussion illustrates the difficulties involved in
assessing the accuracy of DFT, and DFT-based methods, as regards the
calculation of transition temperatures in materials with non-trivial
structural and lattice-dynamical effects like PTO. As mentioned, we
have reasons to think that our result for $T_{\rm C}^{\rm DW}$ might
be smaller than the experimental value, and we also have reasons to
think it might be larger. At any rate, the existence of a DW-confined
polar order and temperature-driven transition remains a solid
prediction.

\section{Relation to previous literature}

As mentioned in our paper, there are several first-principles studies
of the 180$^{\circ}$ DWs of PTO available in the
literature~\cite{poykko99,meyer02,he03,lee09}. Interestingly, except
for the earliest one,\cite{poykko99} all of them suggest that no polar
order occurs within the DW plane. We cannot be sure about the reasons
why these previous works did not find polarized DWs, but some
possibilities are worth mentioning. First, it is usual in studies of
this sort to assume a simplification of the DW structure, typically
imposing that the symmetry operations that are common to the two
neighboring domains be preserved in the structural relaxations. Such
an approximation, which is adopted to alleviate the computational cost
of the simulations, may well render a null DW polarization. Second, we
obtained our polar DW configurations from MC annealings based on our
PTO potential, which allowed us to explore the configuration space
very efficiently, and used such structures as the starting point of
the first-principles relaxations. In our experience, one is not
guaranteed to reach such a distorted structure in a typical
first-principles optimization, which would start from the non-polar DW
state, even if all symmetries are broken. (Note that the information
on the energetics of the DW instability, given in Section~III above,
may be relevant to this issue.)

Let us also note that, recently, Wei {\sl et al}. \cite{wei14} have
found polar distortions at the structural (antiphase) DWs of non-polar
compound PbZrO$_{3}$. Wei {\sl et al}. present first-principles
results showing the bi-stability of the polar DW configuration, and
thus argue that such a polarization is switchable. However, the
supplemental material of Ref.~\onlinecite{wei14} suggests that the
switch of the DW polarization also involves the reversal of structural
distortions {\em inside} the domains. This is reminiscent of the
so-called {\em hybrid improper ferroelectrics} -- where the
polarization switch must be accompanied by the reversal of one
additional order parameter
\cite{bousquet08,benedek11,zanolli13,yang14} -- and of the polar order
predicted at the FS walls of CaTiO$_{3}$
\cite{goncalves-ferreira08}. In contrast, the polarization of our PTO
DWs can switch without affecting the surrounding domains.

%\section*{References}
%\bibliographystyle{unsrt}

%\bibliography{biblio}

\begin{thebibliography}{49}
\bibitem [{fnc()}]{fncell}%
  \BibitemOpen
  \href@noop {} {}\bibinfo {note} {For the FE phase of PTO, our model potential
  predicts long and short lattice constants of 4.21~\AA\ and 3.94~\AA,
  respectively. In contrast, our PBEsol calculations render 4.03~\AA\ and
  3.87~\AA\ for these quantities. These results are consistent with those in
  Figs.~\ref{fig1}(b) and \ref{fig1}(c), as the smaller lattice parameters lead
  to relatively small polar distortions, both in the domains and within the
  DWs, in the PBEsol~(I) case.}\BibitemShut {Stop}%
\bibitem [{fns()}]{fnsymm}%
  \BibitemOpen
  \href@noop {} {}\bibinfo {note} {In this high-$T$ regime, the DW presents all
  the symmetries that are common to its neighboring domains.}\BibitemShut
  {Stop}%
\end{thebibliography}


\end{document}